\documentstyle[multicol,prl,aps,epsf]{revtex}
\begin{document}

\draft

\title{Spin-splitting in GaAs 2D holes}

\author{S. J. Papadakis, E. P. De Poortere, M. Shayegan}
\address{Department of Electrical Engineering, Princeton University,
Princeton, New Jersey  08544, USA.}
\author{R. Winkler}
\address{Institut f\"ur Technische Physik III, Universit\"at
Erlangen-N\"urnberg, Staudtstr. 7, D-91058 Erlangen, Germany.}
\date{\today}
\maketitle

\begin{abstract}
We present quantitative measurements and calculations of the
spin-orbit induced zero-magnetic-field spin-splitting in
two-dimensional (2D) hole systems in modulation-doped GaAs (311)A
quantum wells.  The results show that the splitting is large and
tunable.  In particular, via a combination of back- and front-gate
biases, we can tune the splitting while keeping the 2D hole
density constant.  The data also reveal a surprising result
regarding the magnetoresistance (Shubnikov-de Haas) oscillations
in a 2D system with spin-split energy bands:  the frequencies of
the oscillations are {\it not} simply related to the population of
the spin-subbands.  Next we concentrate on the metallic-like
behavior observed in these 2D holes and its relation to
spin-splitting. The data indicate that the metallic behavior is
more pronounced when two spin-subbands with unequal populations
are occupied.  Our measurements of the magnetoresistance of these
2D hole systems with an in-plane magnetic field corroborate this
conclusion: while the system is metallic at zero magnetic field,
it turns insulating when one of the spin-subbands is depopulated
at high magnetic field.
\end{abstract}

\begin{multicols}{2}
\section{Introduction}
\label{sec:Intro}
Zero-magnetic-field ($B = 0$) spin-splitting of energy bands has
long been a subject of considerable experimental and theoretical
effort because of its role in the electronic properties of
two-dimensional (2D) systems \cite{spinspmore,Lu98,Bychkov+}. It
concerns details of energy band structure that are of fundamental
interest. The eigenstates in a system that has inversion symmetry
in both space and time are spin-degenerate. If the system is made
spatially inversion asymmetric, this degeneracy is lifted by
spin-orbit interaction for all momentum $\vec{k} \neq 0$, even in
the absence of an external magnetic field.  In 2D systems, we can
create this spatial inversion asymmetry simply by making the
potential which confines the carriers to 2D asymmetric.

In this work, we study the $B = 0$ spin-splitting in GaAs 2D holes
confined to a quantum well (QW) grown on the (311)A surface of a
GaAs substrate. Section \ref{sec:exper} provides some experimental
details. In Section \ref{sec:spin}, we show that by using two
gates on the back and front of the sample to control the electric
field applied perpendicular to the QW, the asymmetry and therefore
the spin-splitting can be tuned while the carrier density is kept
constant. In the process we discover that the low-$B$ Shubnikov-de
Haas oscillations are not as simply related to the $B = 0$
spin-splitting as has often been presumed \cite{Winkler00}.

Tuning the spin-splitting without changing the density allows us
to isolate the effect of spin-splitting on various phenomena.  In
recent years there has been significant interest in the surprising
metallic behavior of various high-quality 2D systems
\cite{KravMI,Popovic97,Coleridge97,Lam97,Hanein98,Simmons98,Papadakis98,Papadakis99,Papadakis00,Pudalov97b,Simonian97,Okamoto99,Mertes99,Yoon99b,Murzin98,Yaish99}.
In Section \ref{sec:MIT_spinsp} we show the effect of
spin-splitting on the metallic behavior of the 2D holes in GaAs
\cite{Papadakis99,Papadakis00}.

Finally, motivated by recent theoretical predictions
\cite{Chen99,DasSarmaC99}, we have applied an in-plane $B$ to
these samples and studied the magnetoresistance (Section
\ref{sec:Bpar}).  We have found that, remarkably, the in-plane $B$
depopulates the upper spin-subband at a rate which depends on the
relative orientations of $B$ and the crystal axes! This is a
reflection of the anisotropic band structure and Zeeman splitting
in the system. Furthermore, when the upper spin-subband is
depopulated, the metallic behavior turns to insulating behavior.
We also see an in-plane magnetoresistance anisotropy that depends
on the relative orientations of $B$ and the current $I$; the
finite layer thickness of our 2D system may be responsible for
this observed anisotropy \cite{DasSarmaC99}.

\section{Experimental}
\label{sec:exper}
\begin{figure}
\centerline{\epsfxsize=12cm
\epsfbox{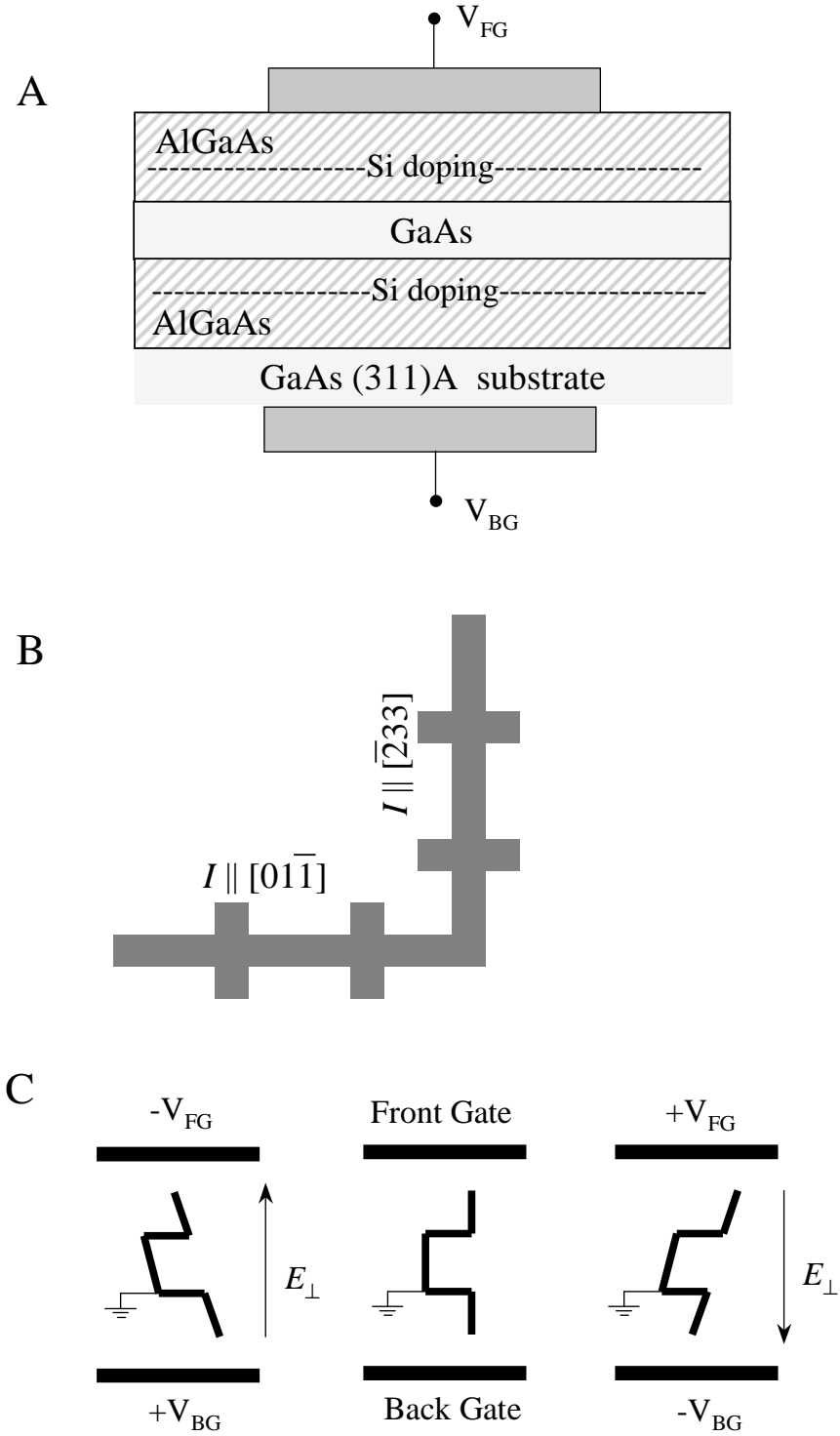}}
\vskip-5pc
\caption{A:  Schematic cross section of the sample, not to scale.
B: Diagram of the $L$-shaped Hall bar used for measuring the
resistivities along the $[01\bar1]$ and $[\bar233]$ directions. C:
Schematic demonstrating how front and back gates can be used to
tune the symmetry of the quantum well, and therefore the
spin-splitting, without changing the density.}
\label{sample}
\end{figure}

A schematic cross section of our samples is shown in Fig.
\ref{sample}A.  The samples are grown, via molecular beam epitaxy
(MBE), on GaAs (311)A substrates.  Each sample contains a 20
nm-wide GaAs QW flanked by AlGaAs barrier layers which are
modulation doped with Si.  On GaAs (311)A substrates, Si is
normally incorporated as an acceptor, leading to a high-quality 2D
hole system in the GaAs QW.  We make ohmic contacts to the 2D
holes using In:Zn and alloying in a reducing atmosphere. As shown
in Fig. \ref{sample}A, our samples have back and front gates to
change the density of the 2D hole system and also tune the
symmetry of the confining QW potential.  The back gate is made by
contacting a layer of In left over on the back of the substrate
from the MBE growth, and the front gate is evaporated metal
(Ti-Au).

GaAs 2D systems grown on (311)A substrates have an intrinsic
mobility anisotropy due to an anisotropic interface roughness
\cite{Heremans94,Wassermeier95}.  The mobility for current along
the $[\bar233]$ direction can be up to several times the mobility
for current along the $[01\bar1]$ direction.  For this reason, we
did our measurements on $L$-shaped Hall bars, which allow
simultaneous measurement of the resistivity $\rho$ for current
along these two directions (Fig. \ref{sample}B).

Measurements were done in dilution and $^3$He refrigerators, with
base temperatures of 30 mK and 0.3 K respectively.  The systems
are fitted in a superconducting magnet capable of magnetic fields
up to 16 T. The $^3$He refrigerator has a tilting stage for the
sample that can be rotated {\it in-situ}, so the relative angle
between the magnetic field and 2D hole system plane can be varied.

Figure \ref{aniso} shows an example of data for one of the samples
from both mobility directions, measured simultaneously at 30 mK .
The inset to A shows the low-$B$ data.  The main part of A
demonstrates the high quality of the sample.  The inset to B shows
raw $T$-dependence of $\rho$ data, and the main part of B shows it
scaled to $\rho_0$, the resistivity at 30 mK.

\section{Tuning Spin-splitting}
\label{sec:spin}

We use the gates to tune the asymmetry of the QW by applying an
electric field ($E_{\perp}$) perpendicular to its plane.  Figure
\ref{sample}C schematically demonstrates the procedure. We set the
front gate ($V_{FG}$) and back gate ($V_{BG}$) voltages, and
measure $\rho$ as a function of $B$. Then, at a small $B$,
$V_{FG}$ is increased and the change in the hole density is
measured from the change in the Hall coefficient. $V_{BG}$ is then
reduced to recover the original density. This procedure changes
$E_{\perp}$ while maintaining the same density to within 3\%, and
allows calculation of the change in $E_{\perp}$ from the way the
gates affect the density.  These steps are repeated until we have
probed the range of $V_{FG}$ and $V_{BG}$ that are accessible
without causing gate leakage \cite{Papadakis99}.  Increasing the
magnitude of $E_{\perp}$ increases the asymmetry of the sample,
which increases the spin-splitting.

Figure \ref{M340B1_33}A shows examples of the Shubnikov-de Haas
(SdH) oscillations measured for current along the $[01\bar1]$
direction in our 20 nm QW, following the procedure described above
to change $E_{\perp}$ from about 5 kV/cm (pointing towards the
front gate) in the top panel through to about -6 kV/cm in the
bottom panel. Beating can be clearly seen in the oscillations of
all traces except the center trace where $E_{\perp} \sim 0$. Fast
Fourier transforms (FFTs) of $\rho$ vs. $B^{-1}$ quantify the
frequencies present. Figure \ref{M340B1_33}B shows the FFTs of the
SdH oscillations at all of the measured sets of $V_{FG}$ and
$V_{BG}$. These frequencies have long been thought to be directly
proportional to the spin-subband densities ($p_{\pm}$) following
\cite{spinspmore,Lu98}:
\begin{equation}
p_{\pm}=\frac{e}{h}f_{SdH}^{\pm}.
\end{equation}
With this interpretation it is clear in Fig. \ref{M340B1_33}B that
we are tuning the spin-splitting through a minimum as we change
$E_{\perp}$ from 5 kV/cm to -6 kV/cm:  the two peaks,
corresponding to two spin-subband densities, get closer together
as $E_{\perp}$ approaches zero, finally merging, only to separate
again as $E_{\perp}$ is decreased away from zero.  Below, we show
that Eq. 1 is only approximately true.

In order to compare the theory of spin-splitting to the
experimental results, we have performed self-consistent subband
calculations that have no adjustable parameters (further details
are in Refs. \cite{Lu98} and \cite{Winkler00}).  These
calculations produce, for a given 2D hole density and $E_{\perp}$,
both the $B = 0$ spin-subband densities and the SdH oscillations.
It is important to note that both of these results are produced by
the same Hamiltonian, so they are directly comparable.  These
calculations yield a surprising result: FFTs of the calculated SdH
oscillations show that the frequencies present, when converted to
spin-subband densities using Eq. 1, do {\it not} agree with the
calculated $B = 0$ spin-subband densities \cite{Winkler00}.

\begin{figure}
\centerline{\epsfxsize=17cm
\epsfbox{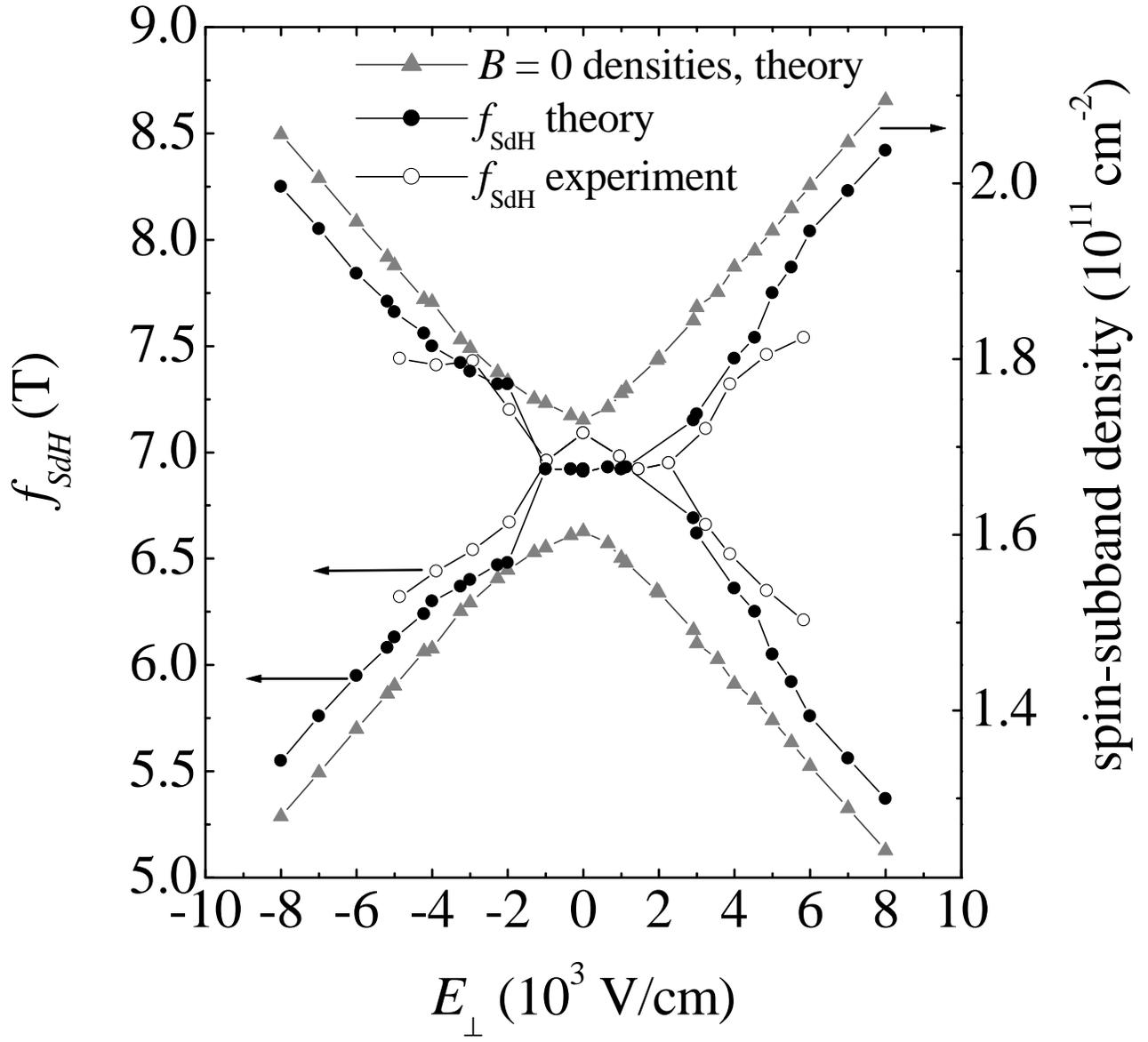}}
\vskip-8pc
\caption{Calculated $B = 0$ spin-subband densities, and the peak positions of FFTs of both the
calculated and measured SdH oscillations.  The left and right axes
are related according to Eq. 1 so the data can be directly
compared.}
\label{Calc_exp}
\end{figure}

This is highlighted in Fig. \ref{Calc_exp} which shows the
calculated $B = 0$ spin-subband densities (right axis), and the
peak positions of FFTs ($f_{SdH}$) of both the calculated and
measured SdH oscillations (left axis), as a function of
$E_{\perp}$ \cite{asym}. The two $y$-axes of the figure are scaled
by Eq. 1, so the sets of data can be directly compared. The
calculated $f_{SdH}$ consistently underestimate the $B = 0$
spin-splitting. The agreement between the theoretical and measured
$f_{SdH}$, however, is very good. Especially noteworthy are the
results for $E_{\perp} \sim 0$.  As the $B = 0$ calculations in
Fig. \ref{Calc_exp} indicate, even at $E_{\perp} = 0$, there
should be finite spin-splitting because of the inversion asymmetry
of the GaAs (zincblende) crystal structure.  However, both the
measured and calculated FFTs show only one peak for $-1 \alt
E_{\perp} \alt 1$ kV/cm \cite{FFTs}.

We believe that the inaccuracy of Eq. 1 is due to a breakdown of
Onsager's  argument \cite{Onsager52}, which is based on
Bohr-Sommerfeld quantization of the semiclassical motion of the
Bloch carriers. The presence of spin-orbit interaction makes the
system inherently quantum-mechanical, so a semiclassical picture
fails.  This finding is quite general in that one can expect
deviations from a semiclassical picture for all systems with
strong spin-orbit interaction. However, the full
quantum-mechanical calculations predict that Eq. 1 can be quite
accurate for some 2D systems grown on high-symmetry crystal
directions, while being inaccurate for other crystal directions or
other systems \cite{Winkler00}.

\section{Metallic behavior and spin-splitting}
\label{sec:MIT_spinsp}

For many years, it was widely accepted that there can be no
metallic phase in a 2D carrier system
\cite{Abrahamsscaling,BishopTsui}.  However, recent experiments on
several different high quality 2D systems have provided us with
reason to re-visit this belief, as they showed that $\rho$ has a
metallic-like temperature dependence: at very low temperatures
$\rho$ decreases with decreasing $T$
\cite{KravMI,Popovic97,Coleridge97,Lam97,Hanein98,Simmons98,Papadakis98,Papadakis99,Papadakis00}.
Various mechanisms
\cite{Pudalov97,Dobrosavljevic97,Altshuler99,DasSarma99} have been
proposed to explain the metallic behavior, but no clear model has
emerged which quantitatively describes the sizeable body of
experimental data.

\begin{figure}
\epsfxsize=17cm
\epsfbox{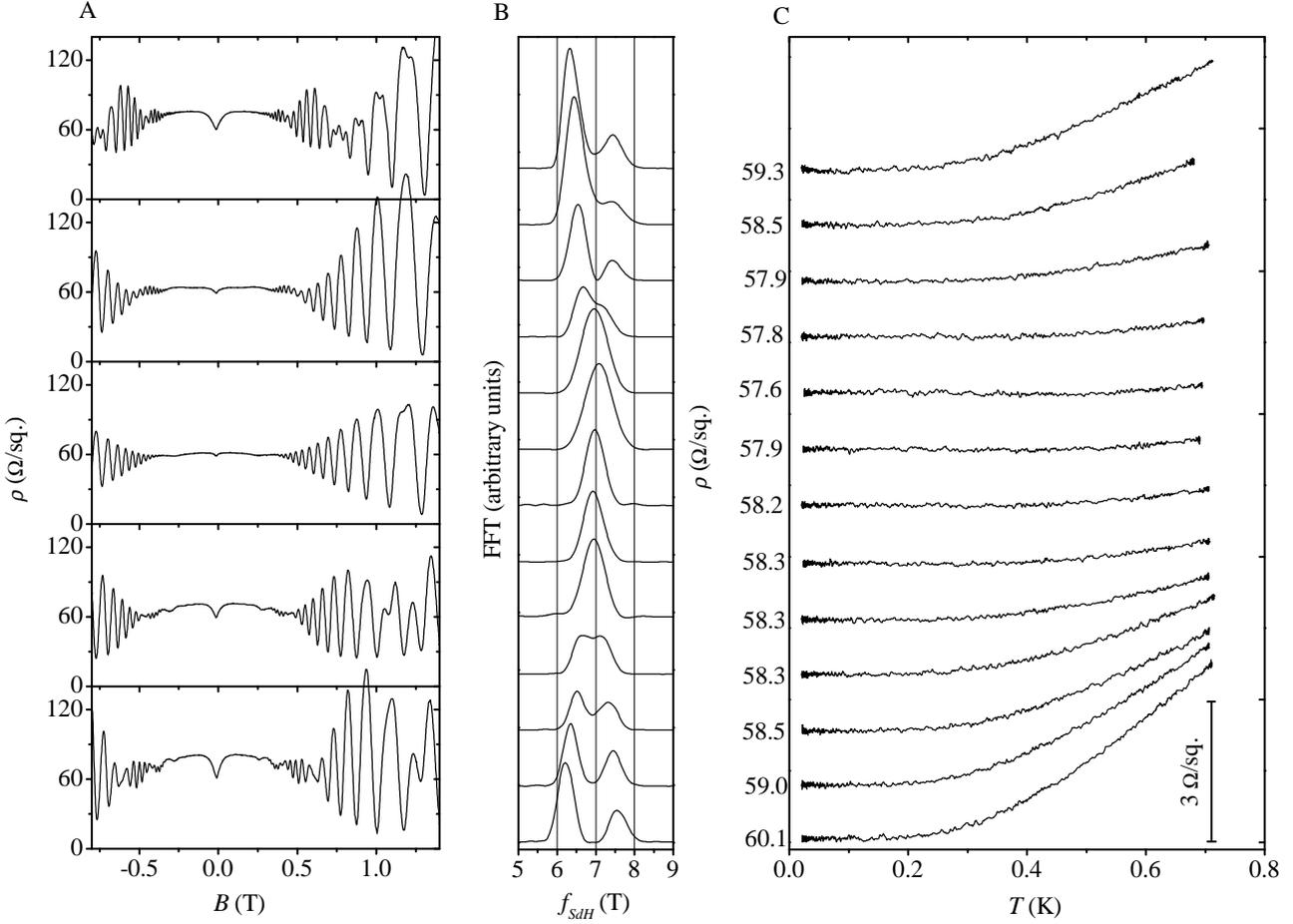}
\caption{A:  Magnetoresistance traces, all at a density of $3.3 \times 10^{11}$
cm$^{-2}$, but at different values of $E_{\perp}$.  The data shown
are from the low-mobility $[01\bar{1}]$ direction. B:  Fourier
transforms of the Shubnikov-de Haas oscillations, showing that the
spin-splitting is being tuned through a minimum.  C:  Temperature
dependence of $\rho$ for the $[01\bar1]$ direction.  The traces
are shifted vertically for clarity, with the value of $\rho$ at 30
mK listed along the $y$-axis for each trace.  Each $\rho$ vs. $T$
trace is aligned with its corresponding Fourier transform.  B and
C together show that the magnitudes of the spin-splitting and the
temperature dependence are related.}
\label{M340B1_33}
\end{figure}

By tuning the spin-splitting at constant density, we isolate the
effect of spin-splitting on the $T$-dependence of $\rho$
\cite{Papadakis99,Papadakis00}. Figure \ref{M340B1_33}C shows
$\rho$ as a function of $T$ for each of the measured $E_{\perp}$.
The traces are lined up with the corresponding FFTs in Fig.
\ref{M340B1_33}B. From this data it is evident that the magnitude
of the change in $\rho$ from 30 mK to $\sim$0.7 K is correlated
with the spin-splitting. The 30 mK, $B = 0$ resistivity ($\rho_0$)
for each trace is listed on the $y$-axis.

\begin{figure}
\epsfxsize=17cm
\epsfbox{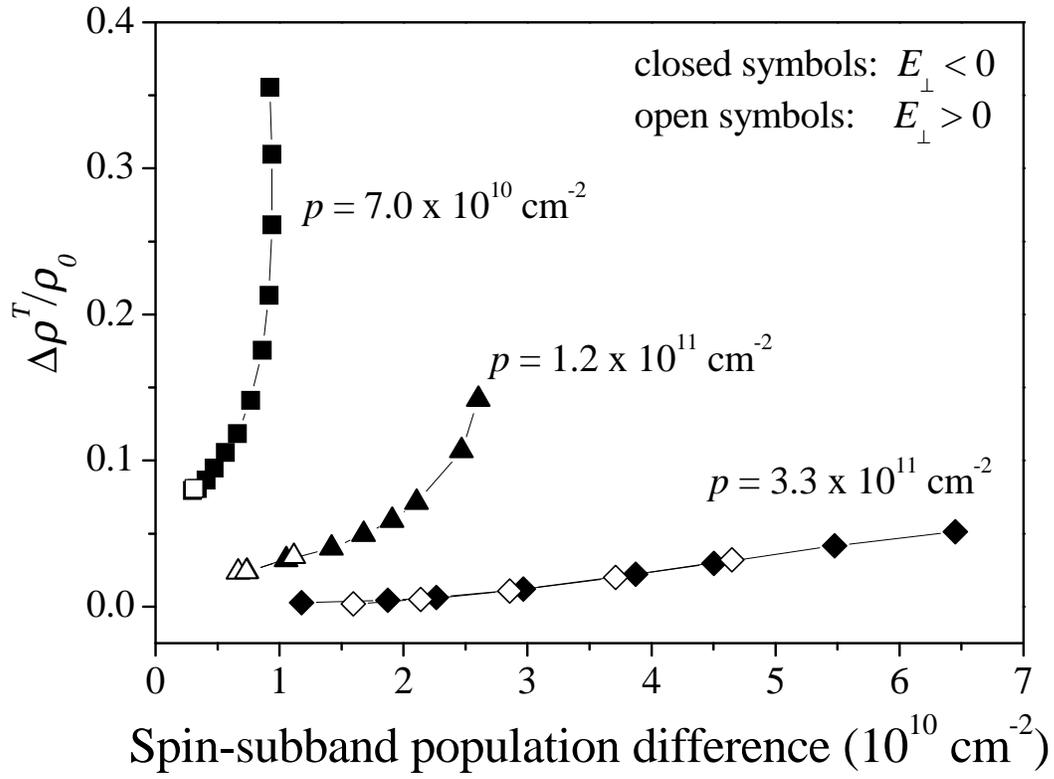}
\vskip-2pc
\caption{Fractional change in resistivity, measured along $[01\bar1]$,
from 30 mK to $\sim$0.7 K plotted vs. the calculated spin-subband
population difference.  It is evident that the magnitude of the
$T$-dependence becomes more sensitive to spin-splitting at lower
densities.}
\label{Tdep_summ}
\end{figure}

In order to characterize the $T$-dependence data in a simple way,
we calculate $\Delta\rho^T/\rho_0$, the fractional change in
resistivity from 30 mK to $\sim$0.7 K.  This is plotted in Fig.
\ref{Tdep_summ} vs. the spin-subband population difference
calculated at $B = 0$ (${\Delta}p_s$).  We have repeated this type
of experiment on lower density samples, down to a density of $2.5
\times 10^{10}$ cm$^{-2}$.  For densities from $7.0 \times
10^{10}$ to $3.3 \times 10^{11}$ cm$^{-2}$, for which the $\rho$
vs. $T$ traces have qualitatively the same shape as in Fig.
\ref{M340B1_33}C, we have calculated the $B = 0$ spin-subband
densities. As shown in Fig. \ref{Tdep_summ}, data for these
densities exhibit the same trend: $\Delta\rho^T/\rho_0$ is larger
at larger $\Delta{p_s}$, and $\Delta\rho^T/\rho_0$ is more
sensitive to spin-splitting at the lower measured densities.

The $T$-dependence of $\rho$ for a density of $2.5 \times 10^{10}$
cm$^{-2}$ has a qualitatively different shape from the
higher-density data: it exhibits a local maximum. Increasing
$\Delta{p_s}$ moves this local maximum to lower $T$.  We note that
previous experiments on the density-dependence of the metallic
behavior show that at high densities $\rho$ monotonically
increases with $T$, and that as the density is reduced and the
transition to insulating behavior is approached, a local maximum
appears in $\rho$ vs. $T$
\cite{KravMI,Hanein98,Simmons98,Papadakis98}. Qualitatively, in
the density range we have measured, the effect of increasing
spin-splitting on shape of $\rho$ vs. $T$ is similar to the effect
of reducing density.

For the highest density, $3.3 \times 10^{11}$ cm$^{-2}$, we have
direct experimental support for the calculated ${\Delta}p_s$.  At
the lower densities the sample quality is typically worse and the
spin-splitting is too small so that two frequencies are not
resolved in the SdH oscillations.   We use the calculations
described in Section \ref{sec:spin} to determine the expected
${\Delta}p_s$ used in Fig. \ref{Tdep_summ} from the $E_{\perp}$.
Even if a large error in ${\Delta}p_s$ is allowed for, the
conclusions of the previous paragraphs are still valid.

Tuning $E_{\perp}$ tunes the spin-splitting, but can also affect
the mobility and cause changes in $\rho_0$.  However, a careful
examination of all our data reveals that the changes in $\rho_0$
are not causing the changes in $\Delta\rho^T/\rho_0$. The
variation of $\Delta\rho^T/\rho_0$ due to $E_{\perp}$ does not
correlate with the changes in $\rho_0$.  One example of this is in
Fig. \ref{M340B1_33}C, where there are three traces at different
$E_{\perp}$ that have different $\Delta\rho^T/\rho_0$, but the
same $\rho_0$ (58.3 $\Omega$/sq.).

All of the data we have presented so far have been from the
low-mobility $[01\bar1]$ arm of the $L$-shaped Hall bar.  The SdH
oscillations in the high-mobility $[\bar233]$ direction data are
very similar to those in the $[01\bar1]$ traces, and the FFTs show
that the frequencies present, as expected, are the same. The
$T$-dependence data, while qualitatively similar, are different
along $[\bar233]$. In a given measurement, $\Delta\rho^T/\rho_0$
is always smaller for the $[\bar233]$ direction than for the
$[01\bar1]$ direction.  An example of this behavior can be seen in
Fig. \ref{aniso}B.

In summary, we have found that in the density regime where $\rho$
increases monotonically with increasing $T$, the magnitude of the
change in $\rho$, $\Delta\rho^T/\rho_0$, is correlated with the
spin-splitting of the 2D hole system.  As the density is reduced,
$\Delta\rho^T/\rho_0$ becomes larger and more sensitive to
increased spin-splitting.  In the density regime where $\rho$ has
a local maximum, increasing the spin-splitting moves the maximum
to lower $T$. We also find that the direction of the current in
the sample plays a surprising role: the higher-mobility direction
shows a smaller $\Delta\rho^T/\rho_0$.

\section{In-plane magnetic field}
\label{sec:Bpar}

We also employed an in-plane magnetic field, using the tilting
stage of the $^3$He refrigerator, to probe the Zeeman splitting in
GaAs 2D holes \cite{Papadakis99d}.  Similar magnetoresistance (MR)
measurements have been recently reported for 2D systems that
exhibit a low-$T$ metallic behavior
\cite{Simmons98,Pudalov97b,Simonian97,Okamoto99,Mertes99,Yoon99b}.
In our measurements we have discovered a remarkable anisotropy in
the effect of $B$ on the spin-subbands, pointing out that the
Zeeman splitting in this 2D system is very anisotropic.

The measurements were done on a sample similar to those described
above, but the densities were lower still. As with previous
samples, we used both front and back gates to control the density.
The longitudinal and Hall MRs were first measured with a
perpendicular $B$, and then the sample was tilted 90$^o$ and
$\rho$ as a function of in-plane $B$ was measured. The
measurements were made once with the sample mounted with the
$[\bar233]$ direction parallel to the tilt axis. Then the sample
was warmed up, re-mounted with the $[01\bar1]$ axis parallel to
the tilt axis, and the measurements were repeated. Note that this
sample also had an $L$-shaped Hall bar geometry as shown in Fig.
\ref{sample}B, so for each orientation of the sample relative to
the in-plane $B$, $\rho$ was measured for both current directions
simultaneously.

In order to make the data easier to track, we scale the traces to
the $B = 0$, base-$T$ value of $\rho$.  Figure \ref{Bparallel}
shows such data for various densities, organized by the relative
orientations of $B$ and the crystal axes.  All traces have an
overall positive MR and, in addition, show a broad feature: there
is an inflection point followed by a reduction in slope, followed
by another inflection point beyond which the traces curve upwards
again.  To highlight this behavior, the arrows in Fig.
\ref{Bparallel} are placed between the two inflection points, at a
$B$ we will refer to as $B^*$. Surprisingly, for each density,
$B^*$ for the $B \parallel [\bar233]$ traces is about 4 T smaller
than for the $B \parallel [01\bar1]$ traces, regardless of the $I$
direction. Also, $B^*$ becomes smaller as the density is reduced.
Figure \ref{Bparallel} reveals that the relative orientations of
$B$ and the crystal axes play an important role in the MR
features.

The existence of the MR features around $B^*$ is intriguing.
Similar, though sharper, features have been observed in in-plane
$B$ measurements in systems with multiple confinement subbands
when a subband is depopulated \cite{Jo93}. Similarly, the features
in our data may be related to a spin-subband depopulation and the
resulting changes in subband mobility and inter-subband scattering
as the in-plane $B$ is increased.  In support of this hypothesis,
numerical calculations similar to the ones described above show
that the spin-subband depopulation happens at much lower $B$ for
$B \parallel [\bar233]$ than for $B
\parallel [01\bar1]$ \cite{Papadakis99d}.  This reflects the
highly anisotropic spin-subband structure of the 2D hole systems
in GaAs (311)A QWs. Our hypothesis is further supported by the
observation (Fig. \ref{Bparallel}) that while $B^*$ clearly
depends on the orientation of $B$ with respect to the crystal
axes, it is rather insensitive to the current direction:
spin-subband depopulation should not depend on the direction of
current in the sample.

At higher in-plane $B$, beyond the MR features around $B^*$, the
data in Fig. \ref{Bparallel} are qualitatively similar.  The
traces for $B \perp I$ have greater slope than the corresponding
traces with $B
\parallel I$, regardless of crystal axes.  In this regime the
magnetic confinement can become comparable to the electric
confinement, and the effects due to the finite-thickness of the
2DHS may be dominant. Indeed, Ref. \cite{DasSarmaC99} predicts
that MR with in-plane $B$ should be significantly larger for $B
\perp I$ than for $B \parallel I$, in agreement with our highest
$B$ data.

\begin{figure}
\epsfxsize=17cm
\epsfbox{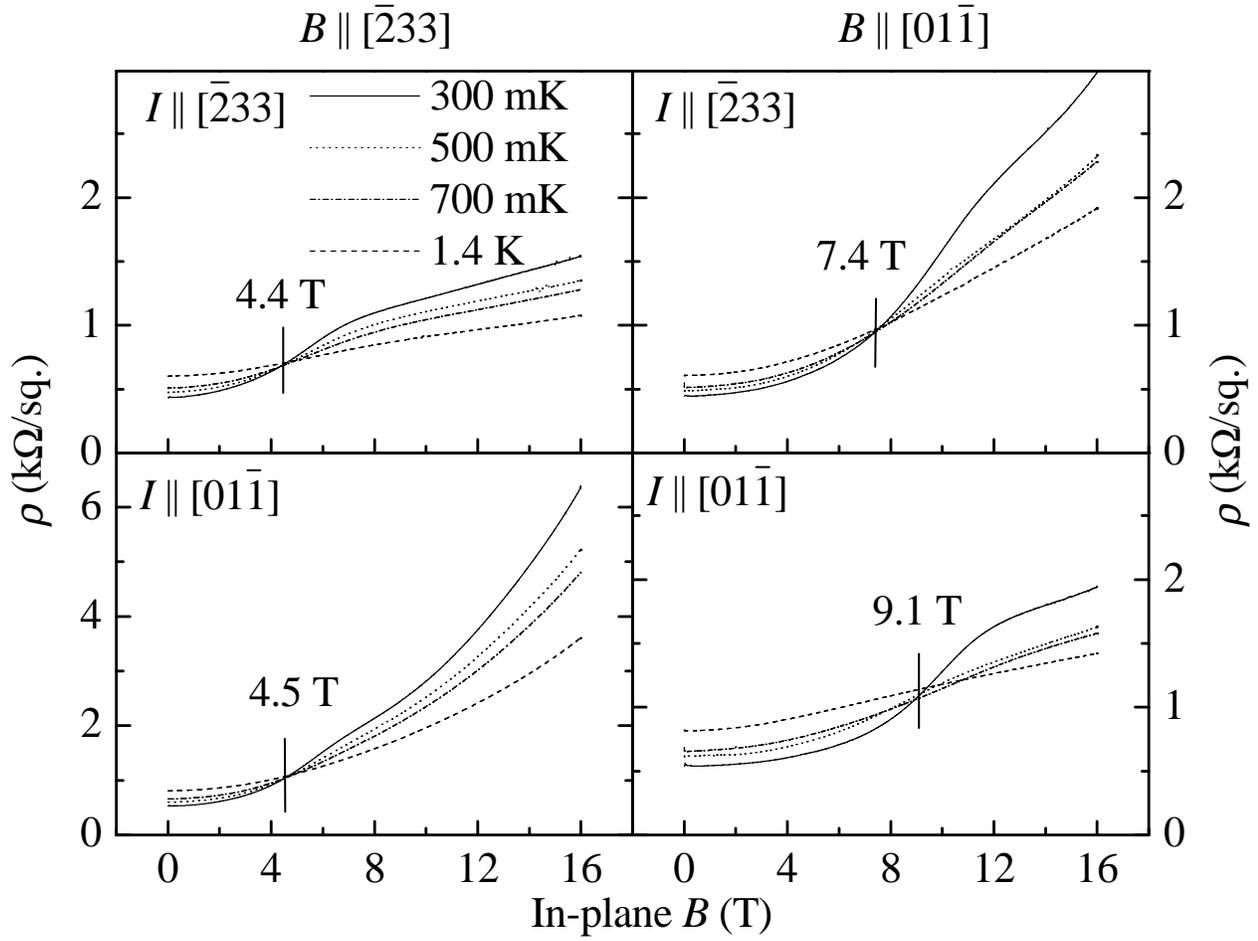}
\caption{Magnetoresistance data at various temperatures, for
density $p = 3.9 \times 10^{10}$ cm$^{-2}$, for the four relative
orientations of $B$, $I$, and crystal axes.  The fields $B_T$ at
which the resistivity is nearly $T$-independent are indicated by
vertical marks.}
\label{temperature}
\end{figure}

Figure \ref{temperature} shows the $T$-dependence of the MR at a
density of $3.9 \times 10^{10}$ cm$^{-2}$, for the four measured
relative orientations of $B$, $I$, and the crystal axes. For each
panel, the traces exhibit a nearly $T$-independent magnetic field
$B_T$ which occurs near the trace's first inflection point. This
is consistent with the data of Ref. \cite{Yoon99b}.  For $B <
B_T$, the data show metallic behavior, and for $B > B_T$,
insulating behavior. $B_T$ is different in each panel and, similar
to $B^*$, depends much more strongly on the orientation of the
crystal axes relative to $B$ than on the orientation of $I$
relative to $B$.  Our experiments indicate that $B^*$ and $B_T$
depend very similarly on the parameters of our systems ($p$,
$E_{\perp}$, direction of $B$).  Our observation, which is in
agreement with the in-plane MR data of Ref. \cite{Okamoto99},
strongly suggests that the metallic behavior is linked to the
presence of two populated spin-subbands
\cite{Papadakis99,Papadakis00,Murzin98,Yaish99}.

\section{Summary}
We have demonstrated tunable $B = 0$ spin-splitting at constant
density in a GaAs 2D hole system.  In the process, we have
discovered that systems with a significant spin-orbit interaction
show a more complicated relationship between the $B = 0$
spin-subband populations and the frequencies present in the
Shubnikov-de Haas oscillations than had previously been expected.
Using the tunability of the spin-splitting to investigate its
effect on the metallic behavior observed in this 2D system, we
find that changing the magnitude of the spin-subband population
difference changes the $T$-dependence of $\rho$.

Through the use of an in-plane magnetic field, we have measured a
surprising anisotropy of the subband structure of (311)A GaAs 2D
holes. When a magnetic field is applied in the plane of the 2D
system parallel to the $[\bar233]$ direction, the upper
spin-subband is depopulated at a significantly lower field than if
the field is applied parallel to the $[01\bar1]$ direction.
Furthermore, we observe that the $B = 0$ metallic behavior turns
into insulating near $B$ at which the upper spin-subband
depopulates.

\begin{figure}
\vskip-1pc
\epsfxsize=16cm
\epsfbox{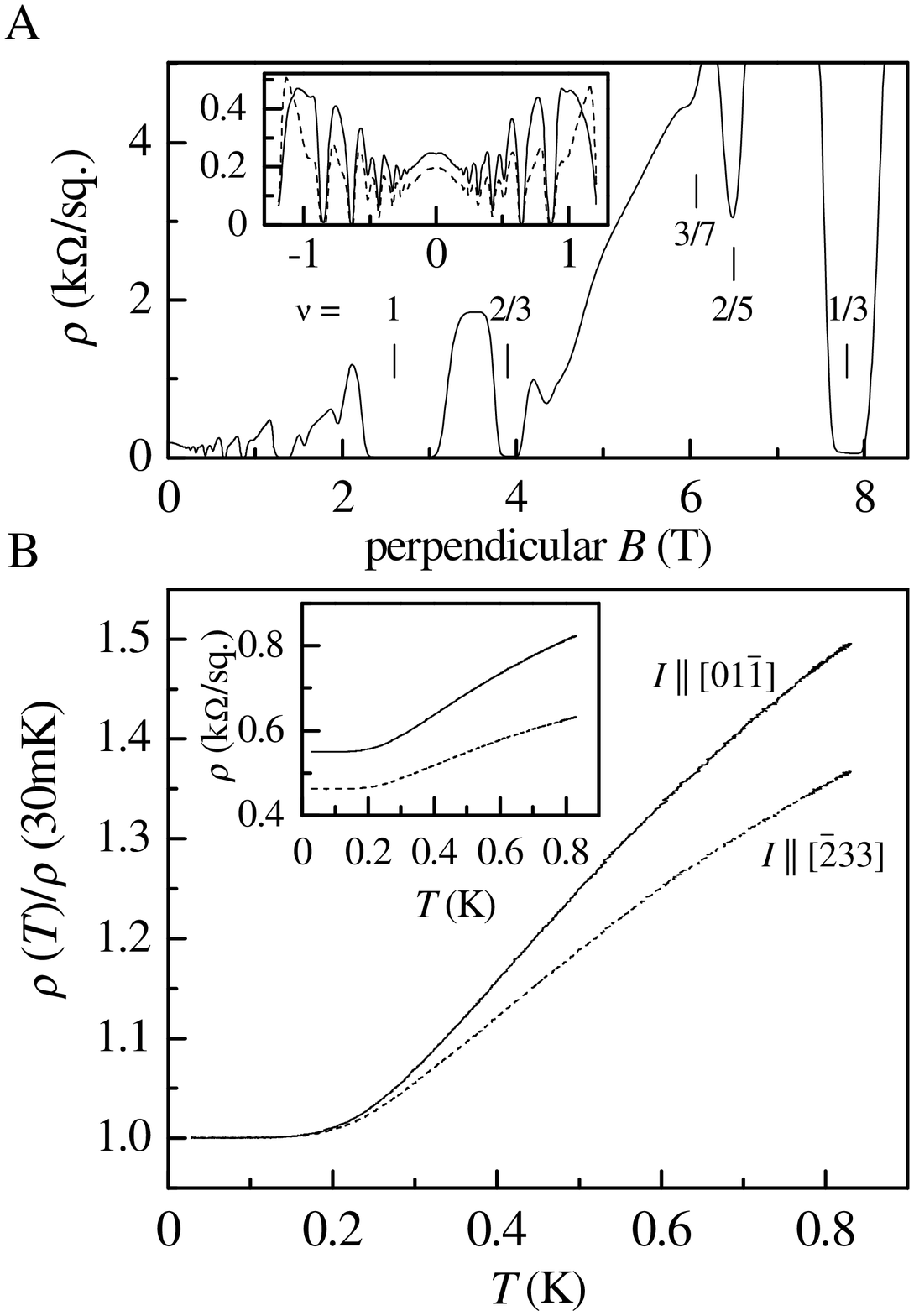}
\vskip-2pc
\caption{A:  Resistivity $\rho$ data for magnetic field $B$
perpendicular to the plane of the 2D hole system with a density $p
=  6.3 \times 10^{10}$ cm$^{-2}$ and current $I
\parallel [\bar233]$ at $T = 30$ mK.  The data exhibits fractional quantum Hall
effect at low filling factors ($\nu$), demonstrating the high
quality of the sample. The inset shows low-$B$ data for $I
\parallel [01\bar1]$ (solid trace) and $I
\parallel [\bar233]$ (dashed trace). B: $B = 0$ temperature-dependence
at $p = 3.3 \times 10^{10}$ cm$^{-2}$, highlighting the difference
between $[01\bar1]$ (solid) and $[\bar233]$ (dashed) directions.
The main figure shows the fractional change in $\rho$ as $T$ is
increased, while the inset shows the raw data.}
\label{aniso}
\end{figure}

\begin{figure}
\epsfxsize=16cm
\epsfbox{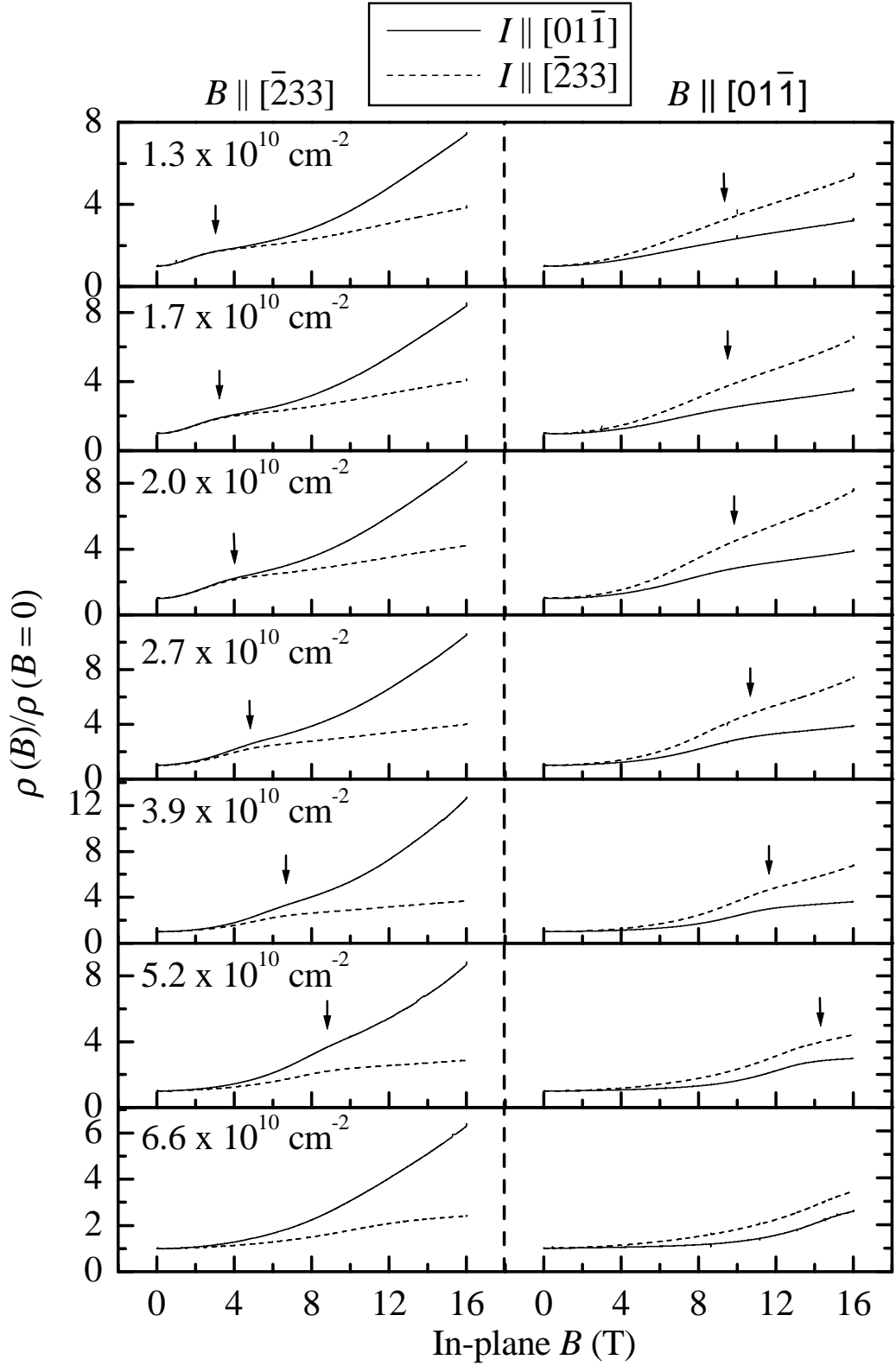}
\caption{Fractional change in resistivity due to an in-plane $B$,
showing that the relative orientations of $B$ and the crystal axes
play an important role in determining the position of the
magnetoresistance features.  The vertical arrows mark $B^*$ as
defined in the text.}
\label{Bparallel}
\end{figure}

Finally, we note that Das Sarma and Hwang have recently reported
calculations aiming to explain the $T$-dependence of the
resistivity \cite{DasSarma99} and the in-plane MR
\cite{DasSarmaC99} of 2D systems that exhibit metallic behavior at
finite $T$.  Their calculations, which include only charged
impurity scattering and the orbital motion, qualitatively
reproduce some of the experimental data.  We wish to point out
that our results reveal the importance of the spin degree of
freedom, and suggest that for an understanding of the experimental
data it is important to also consider a scattering mechanism
involving the spin-subbands, perhaps intersubband scattering
\cite{Murzin98,Yaish99}. Also important for (311)A GaAs 2D holes
is the inclusion of interface roughness scattering:  both the
$T$-dependence of $\rho$ at $B = 0$ (Fig. \ref{aniso}b), as well
as the in-plane MR data (Fig. \ref{Bparallel}), depend on the
direction of the current in the crystal.

This work was supported by the ARO and the NSF.


\begin{thebibliography}{10}
\bibitem{asym}
The asymmetry of the $f_{SdH}$ and the spin-subband densities
visible in Fig.
  \ref{Calc_exp} for $E_{\perp} > 0$ compared with $E_{\perp} < 0$ is due to
  the low-symmetry growth direction [311] of our sample. See Ref.
  \cite{Winkler00} for details.
\bibitem{FFTs}
The resolution of the FFTs, especially the calculated ones, is
good enough that
  they should have displayed two peaks if two frequencies, corresponding to the
  $B = 0$ spin-subband densities predicted by Eq. 1, were present.
\end{thebibliography}

\end{multicols}

\end{document}